\documentclass[pre,aps,twocolumn,floatfix]{revtex4}
\usepackage{bm}
\usepackage{epsfig}
\usepackage{amsmath}

\begin{document}

\newcommand{\ee}{\mathrm{e}}
\newcommand{\ii}{\mathrm{i}}
\newcommand{\dd}{\mathrm{d}}

\newcommand{\tw}{t_\mathrm{w}}

\newcommand{\CC}{\mathcal{C}}

\newcommand{\WW}{\mathbb{W}}

\newcommand{\chinorm}{\chi_4^\mathrm{norm}}

\newcommand{\Ncell}{N_{\rm cell}}

\newcommand{\pf}{c}
\newcommand{\Nf}{N_\mathrm{f}}
\newcommand{\Nm}{N_\mathrm{m}}

\newcommand{\chin}{\chi_\mathrm{4n}}

\title{Dynamical correlations in a glass-former with randomly pinned particles}
\author{Robert L. Jack}
\author{Christopher J. Fullerton} 
\affiliation{Department of Physics, University of Bath, Bath, BA2 7AY}

\begin{abstract}
The effects of randomly pinning particles in a model glass-forming fluid are studied,
with a focus on the dynamically heterogeneous relaxation in the presence of pinning.  
We show how four-point dynamical correlations can be analysed in real space, allowing
direct extraction of a length scale that characterises dynamical heterogeneity.
In the presence of pinning, the relaxation time of the glassy system increases
by up to two decades, but there is almost no increase in either the four-point correlation
length or the strength of the four-point correlations.  We discuss the implications of
these results for theories of the glass transition.
\end{abstract}

\maketitle

\section{Introduction}

Fluids close to their glass transitions exhibit very large relaxation times, as well as characteristic
fluctuations such as intermittent molecular motion and dynamical heterogeneity~\cite{ediger2000,DH-book}.  A central
question in these systems is how these dynamical observations can be related to 
the packing of molecules in space (the liquid structure).  Possibilities include a link
via the diversity of possible packings (the configurational entropy)~\cite{KTW,BB-ktw-2004}, or via
a few privileged structural motifs~\cite{tarjus-review-2005,coslo2007}, or via the emergence of some mobile
`excitations' which are sufficient to determine the dynamical behaviour~\cite{GC-pnas-2003,GC-annrev-2010}.
However, finding meaningful correlations between liquid structure and dynamics is 
challenging, both in simulations and experiments, and demonstrating causal links between structure
and dynamics is harder still.

Recently, this question has been addressed by the procedure of `random pinning'~\cite{berthier-pin2012,Cammarota-pnas2012}.  
A subset of particles is chosen at random and their positions are fixed (quenched) for all future times.  By observing
the motion of the remaining particles in this disordered environment, one aims to elucidate
the relationships between structure and 
dynamics~\cite{kurzidim2011,kim-pin2011,berthier-pin2012,Cammarota-pnas2012,Cammarota-epl2012,karmarkar2012,
jack-plaq2012,charb-prl2012,szamel-pin2013,berthier-pin2013,charb-pin2013}.  
In particular, the prediction of random first
order transition (RFOT) theory~\cite{KTW,BB-ktw-2004} is that the pinning procedure reduces the configurational entropy,
eventually resulting in an ideal glass transition at some special concentration of frozen particles, $c^*$~\cite{Cammarota-pnas2012}.
It is predicted that the transition is accompanied by diverging structural and dynamical length scales that are closely
related to each other.  
If confirmed, these predictions would be strong evidence in favour of RFOT theory.  Recent simulations
have  provided  evidence in favour of a transition at $c^*$~\cite{berthier-pin2013}, based on Monte Carlo
methods that focus on `static' properties of the system (that is, properties of the Boltzmann distribution,
without reference to dynamical motion).  Calculations within mode coupling theory also indicate the
existence of a transition at some $c^*$, although the properties of this transition are not yet fully 
resolved~\cite{krak-mct-transition2011,Cammarota-epl2012,szamel-pin2013}

In this work, we present measurements of dynamical correlations in the presence of random pinning.  
Compared with~\cite{berthier-pin2013}, we use fairly large systems ($1400$ particles), in order to resolve the real-space
behaviour of dynamical (four-point) correlation functions~\cite{DH-book,dasgupta91,benneman99,donati99}.  
The pinning procedure acts to slow down the dynamical relaxation of the 
system -- this limits the ranges of pinning and temperature that we can consider for these system sizes.  However,
we do not find any evidence of an increasing dynamical length scale as random pinning is increased.
Thus, the relaxation time in these systems increases strongly with pinning, at fixed temperature,
without any evidence of increased dynamical heterogeneity or co-operativity.  Of course, these observations
do not rule out the possibility of large static/dynamical length scales at lower temperature
or with more pinning, as predicted by RFOT.  On the other hand, we discuss a simpler theoretical picture based
on an inhomogeneous dynamical arrest, which is sufficient to explain the observations presented here.

We present the models and correlation functions that we will use in Sec.~\ref{sec:model}.
This Section also includes a new method of analysing four-point correlation functions in real space,
to obtain both a dynamic correlation length and the strength of dynamical correlations.  We
present our main numerical results in Section~\ref{sec:results}.  Finally, in Sec.~\ref{sec:discuss}, we discuss
the implications of these results and formulate our conclusions.

\section{Model and observables}
\label{sec:model}

\subsection{Model definition}

We consider the well-studied mixture of Lennard-Jones particles proposed by Kob and Andersen~\cite{ka95}.
It contains two species of particles, A (larger) and B (smaller), in a cubic box
of size $L$, with periodic boundaries.  The natural
unit of length is the diameter of an A-particle, $\sigma=1$, and the energy
unit is the Lennard-Jones energy for AA interactions, $\epsilon=1$.  We
consider a system of $N=1400$ particles, which is large enough to obtain
bulk behaviour for the temperatures shown.
The density is $\rho=1000/(9.4\sigma)^3\approx1.204\sigma^{-3}$ as in~\cite{ka95}.

The system evolves by the Monte Carlo (MC) dynamical scheme described in~\cite{berthier-mc2007}, which gives behaviour consistent
with overdamped Langevin dynamics.  For the time scales associated with structural relaxation,
this MC scheme also gives results that are in quantitative agreement with molecular dynamics.
The natural time unit is $\Delta t = \sigma^2/D_0$ where $D_0$ is the (bare) diffusion constant of
a free particle.  We use the same simulation time step as in \cite{berthier-mc2007}, so that
the time interval $\Delta t$ corresponds to $\frac{3200}{3}\approx 1070$ MC sweeps.  In our numerical
results we set $\Delta t = 1$, which fixes the unit of time.

As in~\cite{berthier-pin2012}, we will consider systems where some particles are pinned in fixed positions,
while all other particles are free to move as usual.  To achieve this, we take an initial
configuration and freeze each particle with probability $\pf$, independently.  Hence, the
number of pinned (frozen) particles $\Nf$ is binomially distributed with mean $\langle \Nf \rangle = \pf N$,
while the number of unpinned (mobile) particles $\Nm$ has mean $\langle \Nm \rangle = (1-\pf)N$.

\subsection{Correlation functions}

When random pinning is present, it is important to focus on ``collective'' overlap functions that depend 
on density profiles but not on particle identities.  To facilitate this, we follow~\cite{berthier-pin2012} and divide
the simulation box into $\Ncell=M^3$ cells of linear size $\ell = L/M$.  We choose the integer
$M=22$ so that $\ell\approx 0.5\sigma$.  
Let $f_i$ be the number of pinned (`frozen') particles in cell $i$,
and $n_i$ be the number of unpinned (`mobile') particles.  The cells are small enough that $n_i+f_i$
is typically zero or unity for each cell.  Similarly, let $n^\mathrm{A}_i$ and $f^\mathrm{A}_i$ be
the number of mobile/frozen particles of type A in cell $i$.


We will use various correlation functions
to describe the system.  
Averages $\langle \cdot \rangle$ run over the configurations of the system and over the choices
of which particles are pinned.
The average cell occupancies are $\langle n_i \rangle = \langle \Nm \rangle/\Ncell$ 
and $\langle f_i \rangle = \langle \Nf \rangle/\Ncell$. 
In the following, 
our correlation functions focus on particles of type A, but we emphasize that when randomly
pinning particles, we treat types A and B equally.
To measure static correlations associated with liquid structure,
we define
\begin{equation}
g_{ij} = \frac{ \langle n^\mathrm{A}_i (n^\mathrm{A}_{j}- \delta_{ij}) \rangle }{ \langle n^\mathrm{A}_i \rangle^2} .
\end{equation}
By translational invariance, $g_{ij}$ depends only on the relative position of cells $i$ and $j$.
We therefore denote the spherical average of this function simply by $g(r)$.  
This $g(r)$ mirrors the behaviour of the familiar radial
distribution function of the fluid.
We also define a two-time collective overlap function
\begin{equation}
F(t,t') = \frac{ \langle n^\mathrm{A}_i(t) n^\mathrm{A}_i(t')\rangle - \langle n^\mathrm{A}_i \rangle^2 }{ \langle (n^\mathrm{A}_i)^2 \rangle - \langle n^\mathrm{A}_i \rangle^2},
\end{equation}
which decays from $1$ to zero as the system's density profile decorrelates.  In 
the presence of pinned particles, the system never entirely decorrelates from its initial
state so $q \equiv \lim_{t\to\infty} F(t)>0$ if $\pf>0$.

To measure dynamical heterogeneity in the system, we define the dynamical overlap for cell $i$:
\begin{equation}
a_i(t,t') = n^\mathrm{A}_i(t) n^\mathrm{A}_i(t'),
\end{equation}
noting that $F(t,t')$ is proportional to $\langle a_i(t,t') \rangle - \langle n_i^{\rm A}\rangle^2$.
We then consider `four-point' correlation functions~\cite{DH-book,dasgupta91,benneman99,donati99} that are constructed from the two-point correlations of the dynamical overlap:
\begin{equation}
g_{4,ij}(t,t') = \frac{ \langle a_i(t,t') [a_{j}(t,t') - \delta_{ij}] \rangle }{ \langle a_i(t,t') \rangle^2 } .
\end{equation}
Assuming time-translation invariance, we may set $t'=0$ without loss of generality,
and we take a spherical average of this function to obtain $g_4(r,t)$, by analogy with $g(r)$.  
For large $r$, we have $g_4(r,t)\to 1$.  For $t=t'$ we have $g_4(r,0)\approx g(r)$;
if the configurations at $t$ and $t'$ are uncorrelated then 
$g_{4,ij}(t) = g_{ij}^2$, in which case $g_4(r,t)$ is very well-approximated by $g(r)^2$.

Finally, we define the global overlap $Q(t,t') = \sum_i a_i(t,t')$ and
a four-point susceptibility $\chi_4(t,t')$ through
\begin{align}
\chi_4(t,t') & = \frac{1}{N_{\rm A}} \big[ \langle Q(t,t')^2 \rangle - \langle Q(t,t') \rangle^2 \big] .
\end{align}
We note that the cell size $\ell$ plays the part of the `probe' length scale in $\chi_4$~\cite{DH-book,chandler06}.

In order to obtain an accurate measure of the extent of dynamical
heterogeneity, it is useful to normalise $\chi_4$ by a time-dependent factor:
\begin{equation}
\chin(t,t') =  \chi_4(t,t') \cdot \frac{\langle n_i^{\rm A}\rangle^2}{\langle a_i(t,t') \rangle^2}
\label{equ:def-chin}
\end{equation}
where the subscript `n' indicates that the function has been normalised.
We note that with this choice
\begin{equation}
\chin(t,t') =   \langle n_i^\mathrm{A}\rangle \sum_{j} [g_{4,ij}(t,t')-1]  + 
\frac{\langle n_i^\mathrm{A}\rangle}{\langle a_i(t,t') \rangle} 
\label{equ:chin-g4}
\end{equation}
The dominant behaviour in $\chin$ comes from the first term, which indicates the extent of spatial correlation
among cells: the second term is a rather trivial `self' contribution.  The time-dependence of the normalisation 
in (\ref{equ:def-chin}) is motivated by 
the observation that $g_{4,ij}(t,t')$  measures the (dimensionless)
relative enhancement of the probability that $a_j(t,t')=1$, given that $a_i(t,t')=1$.  Since $\chin$
can be expressed as a sum over $g_{4,ij}$ without any time-dependent prefactors, we argue that this quantity
is the most appropriate choice when measuring the strength and range of four-point dynamical correlations.
The leading prefactor of $\langle n_i^\mathrm{A} \rangle$ is motived by the observation that for small enough cells,
one has $\langle n_i \rangle \sum_j (\cdot) \to \rho \int\mathrm{d}^3\bm{r} (\cdot)$ where $\rho=N/V$ is the particle
density.  This allows a connection between the cell observables defined here and the more familiar density fields
of liquid state theory.

%

\begin{figure}
\includegraphics[width=6cm]{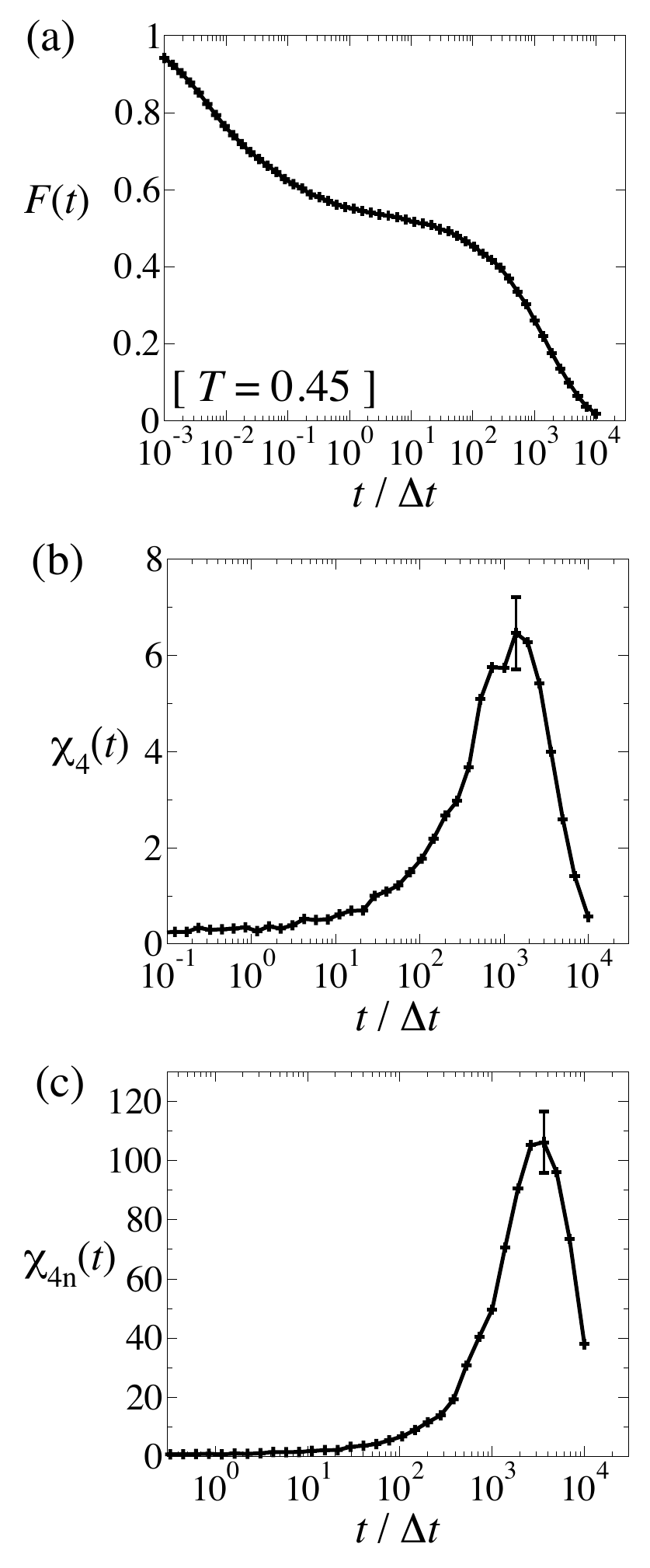}
\caption{Dynamical relaxation and four-point susceptibilities at temperature $T=0.45$, without pinning.
Illustrative error bars were estimated by resampling.
}
\label{fig:T045-ct}
\end{figure}

\begin{figure}
\includegraphics[width=6cm]{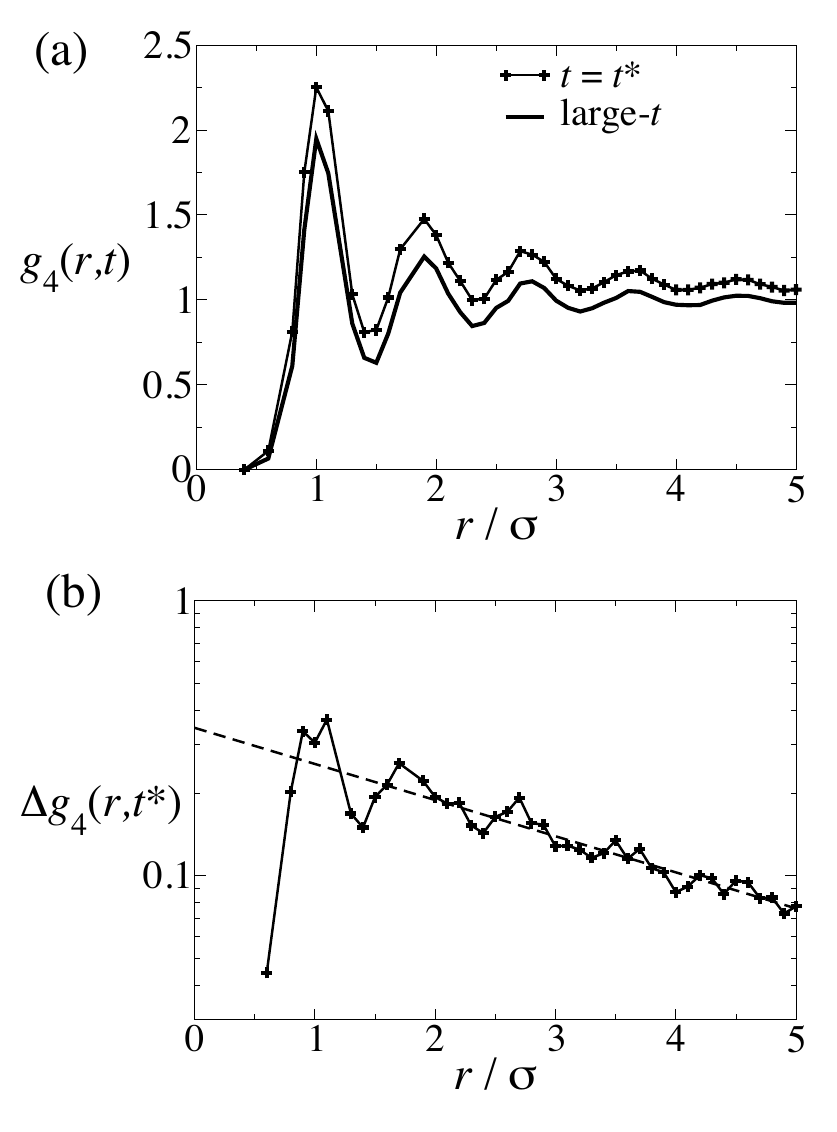}
\caption{Four-point correlation functions in real space, at $T=0.45$, without pinning.
(a) Comparison of $g_4(r,t\to\infty) $ with $g_4(r,t^*)$, where $t^*$ is the time that maximises $\chin(t)$.
(b)~Difference $\Delta g(r,t^*)$,
showing the exponential decay of spatial correlations.  The dashed line is a fit over the range $2\sigma\leq r\leq 5\sigma$ to the form
 $\Delta g(r,t^*) = A\ee^{-r/\xi}$ with
$A=0.34$ and $\xi=3.3\sigma$.
}
\label{fig:T045-g4}
\end{figure}

\begin{figure*}
\includegraphics[width=11cm]{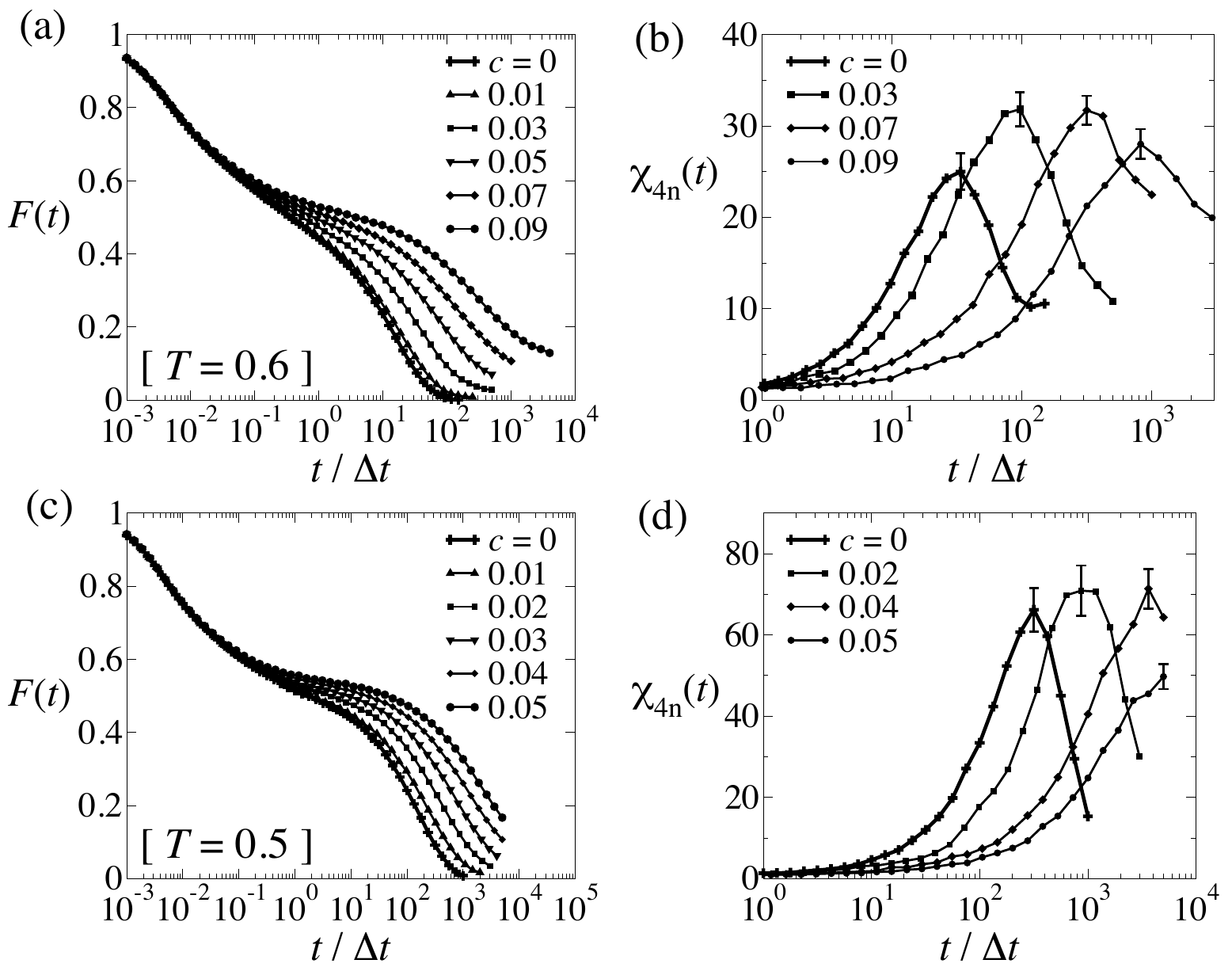}
\caption{
  Correlation functions in the presence of random pinning. (a,c)~The cell autocorrelation
  function $F(t)$ at temperatures $T=0.6$~(in a) and $0.5$~(in c), for varying concentrations $c$ of pinned particles.
  (b,d) The normalised susceptibility $\chin(t)$ at $T=0.6$~(in b) and $0.5$~(in d), for illustrative values of $c$.  The
   maximal value of $\chin(t)$ varies weakly with $c$, despite a significant increase in relaxation time.
}
\label{fig:dyn}
\end{figure*}

\subsection{Illustrative results, without pinning}

To illustrate the behaviour of the various correlation functions, we present data from equilibrium
simulations (without pinning) at the fairly low temperature $T=0.45$.  The system is time-translation
invariant so we set $t'=0$ and suppress any dependence of our correlation functions on this time, 
for compactness of notation.

Fig.~\ref{fig:T045-ct} shows $F(t)$ and the dynamical susceptibilities $\chi_4(t)$ and $\chin(t)$.
Both $\chi_4(t)$ and $\chin(t)$
show the usual peak near the structural relaxation time.  
However, since the time-dependent normalisation factor in (\ref{equ:def-chin}) decreases with time, the peak
in $\chin(t)$ occurs at a later time than the peak in $\chi_4(t)$.  
Fig.~\ref{fig:T045-g4}(a) shows $g_4(r,t)$, together with its long time limit $g(r)^2$.  These data prompt us to consider
%
\begin{equation}
\Delta g_4(r,t) = g_4(r,t) - g(r)^2
\end{equation}
which is shown in Fig.~\ref{fig:T045-g4}(b), for $t=t^*$, the time that maximises $\chin(t)$.

There is some structure in $\Delta g_4(r,t)$ on short length scales that mirrors the oscillations in $g(r)$,
but the behavior on longer length scales is well-described by a single exponential.  
By fitting this exponential decay, we can therefore estimate the
length scale associated with four-point correlations as $\xi_4 = 3.3\sigma$, broadly consistent with earlier studies~\cite{toninelli05,berthier-chi4-07,DH-book}.
We note that $\Delta g_4(r,t)\to0$ as $t\to\infty$ in the absence of pinning,
and this measurement is based on subtracting the long time limit of $g_4(r,t)$ from its value at time $t$.
This procedure yields a function that is (roughly) monotonic, which aids fitting;
it is made possible by the `collective' four point functions considered here. It is more
common to consider four-point functions based on `self'-correlation functions for which the long
time limit of $g_4(r,t)$ is hard to estimate numerically, making it difficult to define an
appropriate analog for $\Delta g_4(r,t)$.

The single exponential decay in $\Delta g_4(r,t^*)$ is expected if the dynamical heterogeneities are clusters of mobile/immobile particles that
are `compact': that is, their `fractal dimension' appears to be close to
three, equal to the spatial dimension~\cite{DH-book}.  
Other fits, for example to the Ornstein-Zernicke form $\Delta g_4(r,t^*)=A\ee^{-r/\xi}/r$, are also possible~\cite{szamel11}, but the single
exponential shown does give a strikingly good fit.
The presence of a single exponential also leads to a contribution to the
four-point structure factor~\cite{benneman99,donati99} given (for small-$q$)by
$S_4(q,t^*) \sim \frac{\xi_4^{-1}}{(q^2+\xi_4^{-2})^2} \sim \frac{\xi_4^3}{1+2q^2\xi_4^2}+O(q^4)$.
[Here $S_4(q,t)$ is simply the (three-dimensional) Fourier transform of $g_4(r,t)$.]  We note however that the
real-space fit in Fig.~\ref{fig:T045-g4} (and Fig.~\ref{fig:g4} below) leads directly to a length scale
and avoids some of the subtleties associated
with fits to $S_4(q,t)$~\cite{karmarkar09,szamel10} (see also~\cite{szamel11}).

We also note that while $\chi_4(t)$ depends on the choice of ensemble used in computer 
simulations~\cite{DH-book,berthier-science05,berthier-chi4-07} 
the function $g_4(r,t)$ depends on the choice of ensemble only through finite-size corrections at
 $O(1/N)$.  [These corrections do not decay with $r$ and therefore 
yield a finite contribution after the sum in (\ref{fig:T045-g4}), resulting in an 
ensemble-dependent $\chi_4(t)$.]  Further, the energy is free to fluctuate in our
MC simulations, which helps to minimise finite-size corrections to $g_4$. When we introduce random pinning then the number of
pinned particles is also free to fluctuate, again minimising these finite-size corrections.  Some further discussion
of this point is given in the Appendix.


\section{Numerical results with pinned particles}
\label{sec:results}

\begin{figure*}
\includegraphics[width=12cm]{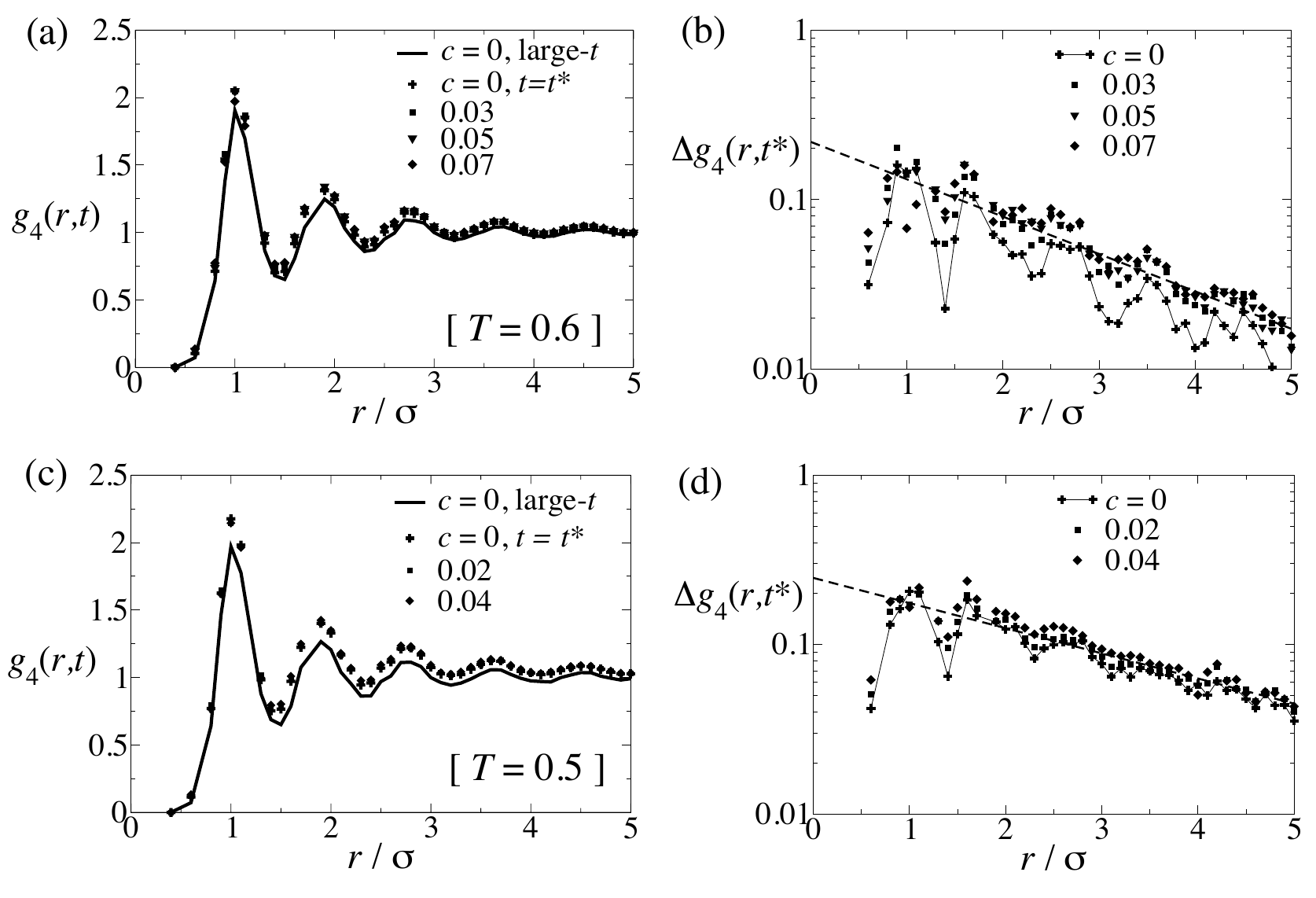}
\caption{Real-space correlations for dynamical heterogeneity in systems with pinning.  (a,c)~The
four-point function $g_4(r,t)$ at $T=0.6$~(in~a) and $0.5$~(in~c).  Results for $c>0$ are taken at $t=t^*$, and compared with
results for $c=0$ at $t=t^*$ and $t\to\infty$.  (b,d)~The difference
$\Delta g_4(r,t^*) = g_4(r,t^*) - g(r)^2$ showing single exponential decay, at $T=0.6$~(in~b) and $0.5$~(in~d).  
The dashed lines are fits to 
$\Delta g(r,t^*) = A\ee^{-r/\xi}$ with
$(A,\xi)=(0.22,2.0\sigma)$ for $T=0.6$ (panel b) and $(0.25,2.9\sigma)$ for $T=0.5$ (panel d).}
\label{fig:g4}
\end{figure*}

We now turn to the main question of this article: how does the presence of pinned particles affect
the dynamical relaxation of the system. 
Figs.~\ref{fig:dyn} and \ref{fig:g4} show results obtained at temperatures $T=0.6$ and $T= 0.5$ 
using the pinning protocol of~\cite{Cammarota-pnas2012}.  That is, starting from an equilibrium configuration, each particle
is pinned with probability $\pf$, independently of other particles.  The pinning takes place at time $t'=0$ and
we measure correlation functions between this time and $t$.  As before, the ensemble is time-translation invariant
so we suppress the dependence of correlation functions on $t'$, for compactness of notation.

Figs.~\ref{fig:dyn}(a,c) show that the relaxation time increases by up to two orders of magnitude for the ranges of
$\pf$ shown.  In the presence of pinning, the shape of the function $F(t)$ also appears more `stretched' (non-exponential), indicating
the presence of a broad range of time scales. The presence of a clear plateau in $F(t)$ indicates that the system
is behaving in a glassy fashion, and one expects co-operative and heterogeneous dynamics to be taking place.
However, the large time scales encountered on pinning limit the range of temperature
and $\pf$ considered: these are temperatures at which the system is moderately supercooled and $\pf$ is small enough that
the long time limit of $F(t)$ is closer to zero than it is to the plateau.  

The key result of Figs.~\ref{fig:dyn}(b,d) is
that the peak of the normalised susceptibility $\chin(t)$ increases slightly for small $\pf$ but quickly saturates at a value
that is almost independent of $\pf$.  In all cases, $\chin(t)$ continues to show a clear peak
as a function of time: we note that for larger $\pf$ when the system does not decorrelate significantly even as $t\to\infty$, we expect
$\chin(t)$ to become monotonic, with a large-$t$ limit whose value indicates 
the heterogeneity of the response to pinning.  However, we do not observe this effect, since increasing $\pf$ further makes the dynamics
very slow, and prevents sampling in the long-time limit.

In Fig.~\ref{fig:g4}, we show $g_4(r,t^*)$, where $t^*$ is the time at which $\chin(t)$ is maximal.  The data show 
clearly that there is no significant change in the four-point correlation
length $\xi_4$, as pinning is increased.  We again emphasise that $g_4(r,t)$ is a dimensionless function of order unity, so
that both the strength of the correlations and their range have direct physical interpretation.  Hence the fact
that $g_4(r,t^*)$ changes little as $c$ is increased is indicating that the dynamical heterogeneity responds weakly to pinning.
Since $\chin(t^*)$ is obtained by a spatial integral of $[g_4(r,t^*)-1]$, the weak dependence of $\chin(t^*)$ on $c$
in Fig.~\ref{fig:dyn}(b,d) is directly attributable to the weak dependence of the dynamical heterogeneities on $c$.


Finally, in Fig.~\ref{fig:chin-all}, we combine the various measurements of $\chin(t)$, evaluating
them at $t^*$ for the various state points considered.  It is clear that if we compare state points according
to their relaxation times, then $\chin(t^*)$ depends
strongly on the temperature, but weakly on the concentration of pinned particles.  
We have emphasised the normalisation of $\chin(t)$ in this article, since we are using $\chin(t)$ in Fig.~\ref{fig:chin-all}
as a proxy for the dynamical length scale $\xi_4$: further evidence that $\xi_4$ depends weakly on $c$ is given in Fig.~\ref{fig:g4}.
The Appendix shows some results for alternative dynamical susceptibilities, which do increase with $c$ (at constant temperature). 
However, in contrast to the transparent relation between $\chin(t)$ and $\xi_4$, 
the increasing values of these alternative susceptibilities do not appear to be directly attributable to increasing length scales.
In the remainder
of this article, we discuss the numerical results of Figs.~\ref{fig:dyn}-\ref{fig:chin-all} 
in the light of the theory of random pinning and of the glass transition.

\section{Discussion}
\label{sec:discuss}

The recent interest in effects of random pinning on glassy systems have focussed on the proposal~\cite{Cammarota-pnas2012}
that an ideal glass transition should occur on pinning, at some $\pf^*$.  This transition has the flavor
of a critical point, in that the relaxation time diverges, as do
length scales (including $\xi_4$) and susceptibilities (like $\chin$). 
Both random first-order transition (RFOT) theory and mode coupling theory (MCT) make predictions that
indicate how these quantities should increase with $c$~\cite{Cammarota-epl2012}.  
At low temperatures and relatively small $c$ (not too close to $\pf^*$), one expects (from MCT) a power-law
relationship between $\chin$ and the relaxation time $\tau$.  This power law then crosses over to a weaker (logarithmic)
dependence (from RFOT) as the critical $\pf^*$ is approached.  On the other hand, at higher temperatures, there is no
transition at any $c^*$.  The highest temperature for which the transition exists is called $T_h$~\cite{Cammarota-pnas2012,
Cammarota-epl2012}, where one expects (within MCT) a logarithmic dependence of $\chi_4$ on $\tau$~\cite{biroli-a3}.

\begin{figure}
\includegraphics[width=6.5cm]{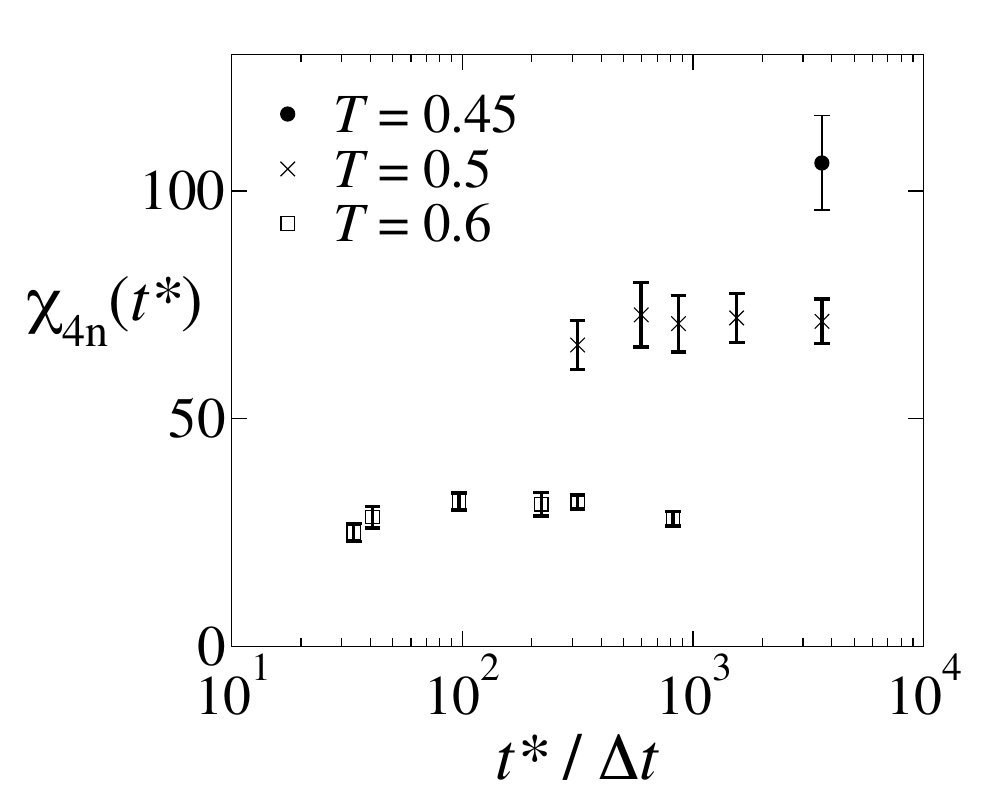}
\caption{The maximal susceptibilities $\chin(t^*)$, plotted against the times $t^*$ at which
these maximal values occur.  The data are shown for $0\leq \pf\leq 0.09$ at $T=0.6$; for $0\leq \pf\leq 0.04$ at $T=0.5$,
and for $\pf=0$ at $T=0.45$}
\label{fig:chin-all}
\end{figure}

For the state points considered here,
we have found that the relaxation time increases strongly, without the increases in $\xi_4$ or $\chin$ that are predicted
by RFOT/MCT.  One would
expect even a relatively weak logarithmic dependence of $\chin$ on $\tau$ to be apparent as an increase in Fig.~\ref{fig:chin-all}.
However, there is
no direct contradiction between our results and the predictions of~\cite{Cammarota-pnas2012,Cammarota-epl2012}, 
since the temperatures considered here are not very 
low, and the values of $\pf$ not very large (in particular, one may have $T_h<0.5$).
One might find an increasing susceptibility (and length scale) 
at lower temperature or larger $c$.  
However, we have not found any concrete evidence in favour of a random pinning glass transition in these
dynamical measurements.

In terms of the random first-order transition (RFOT) theory,
another relevant consideration is the size of the point-to-set length 
scale~\cite{BB-ktw-2004,jack-pts05,montanari06,cavagna08,berthier-pin2012} in the presence of pinning $\xi_{\rm PTS}(c)$,
and its interplay with the four-point length scale $\xi_4$.  
In the presence of pinned particles, the point-to-set length should be obtained by choosing a region in the system of size $R$,
and pinning all particles outside that region, as well as the pinned particles already contained inside the region.  For
small $R$, the remaining mobile particles are unable to relax and the system can access only a single state, while for large $R$, particles
inside the cavity are unaffected by the frozen particles outside, so the system can access many states.  The point-to-set
length is (loosely speaking) the value of $R$ that marks the boundary between the single-state and the many-state regimes.
Structural relaxation in the system requires correlated motion on a scale at least as big as $\xi_{\rm PTS}(\pf)$, 
so one expects $\xi_4 \geq \xi_{\rm PTS}(\pf)$ in general.  Thus,
the four-point lengths estimated here represent upper bounds on $\xi_{\rm PTS}(\pf)$, indicating that it
has not grown above $3\sigma$ for the range of $\pf$ considered.  If $\xi_{\rm PTS}(\pf)$ 
grows strongly with $\pf$ as $c^*$ is approached, one should observe a rapid increase in $\xi_4$
and $\chin(t^*)$ as $c$ approaches $c^*$.

Since we have not found direct evidence in favour of this RFOT scenario, it is useful to consider
an alternative picture of the response to pinning.  Clearly, pinning particles in the system
reduces the available motion of the remaining particles, and acts to slow down structural relaxation.
Further, the existence of dynamical heterogeneity in the system indicates that the response of
dynamical properties to pinning should also be heterogeneous.  For example, if a certain
region of the system has a relatively low propensity for motion~\cite{asaph04,asaph07}, one can expect that a relatively
low number of pinned particles should be sufficient to prevent relaxation in that region.  On the
other hand, regions with higher propensity will be less affected by pinning.  If one simply assumes that
these regions are independent, the result is a smooth crossover from relaxational behaviour at small $\pf$
to arrested behavior at larger $\pf$, and no increase in $\chin(t^*)$ would be observed.  
(This assumes that the dynamical length scale measured by $\chin(t^*)$ is similar to that associated with
high/low propensity regions, consistent with~\cite{berthier-predict07}.)  

We emphasise that this picture of different regions of the system responding almost independently to pinning is
an opposite extreme to the mean-field treatment of RFOT, where correlations between well-separated
regions of the system appear, as in classical phase transitions.  The question of whether neighbouring
regions are coupled seems to be the crucial one in evaluating how systems respond to pinning.  The 
picture of weakly-correlated domains described here is consistent with our results and it is also
broadly consistent with the behaviour in the spin model considered in~\cite{jack-plaq2012}. There is a mild
increase in $\chin(t^*)$ with $\pf$ in that system, 
presumably due to weak correlations between different regions, as they respond to the pinning.  

However, the idea of weakly-correlated domains in supercooled liquids
appears to be at odds with recent results~\cite{berthier-pin2013} showing
that systems of up to 128 particles respond to pinning by a sharp transition from a fluid to an arrested state.
This result is in accordance with the predictions of RFOT. 
However, it would also be consistent with the weakly-correlated domain picture if the size of the relevant domains were
comparable with the size of the system considered.  An analysis of the interplay between four-point dynamical
lengths and the static measurements of~\cite{berthier-pin2013} might help to resolve this question. 

\begin{figure}
\includegraphics[width=6.5cm]{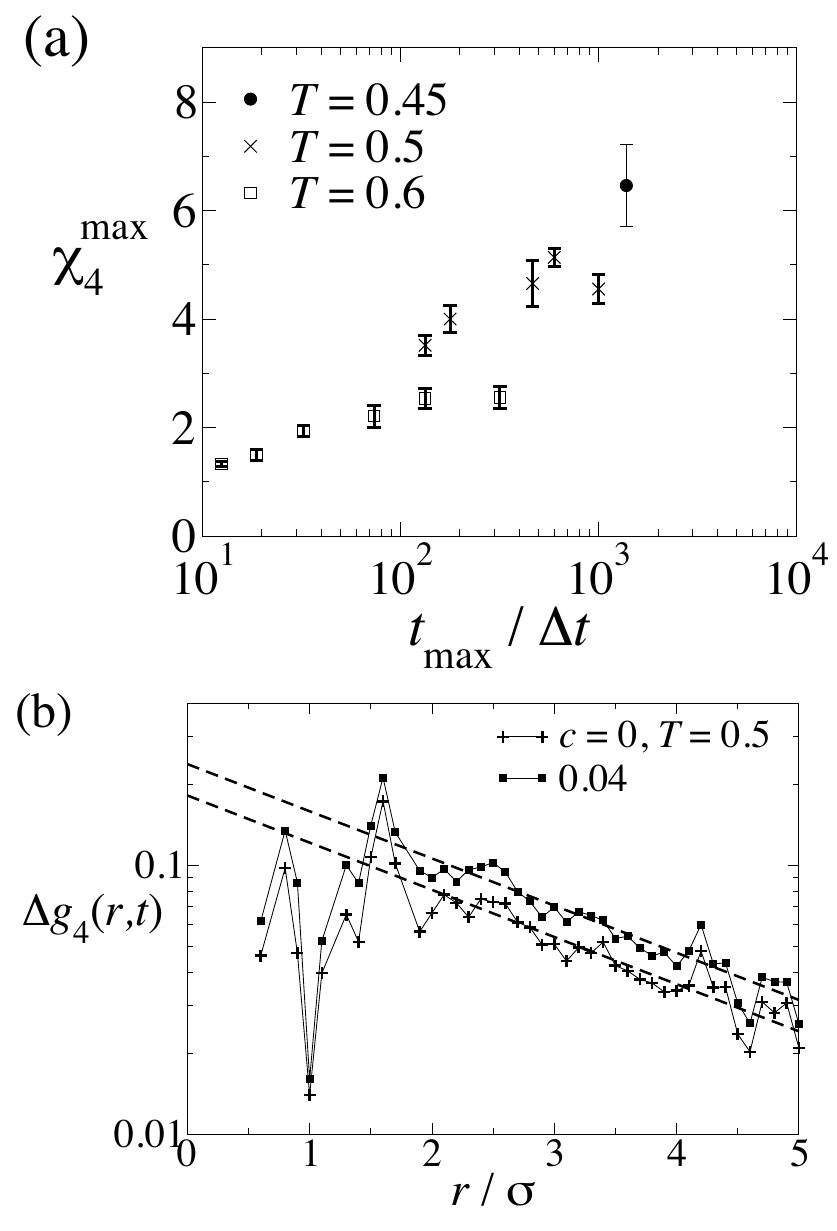}
\caption{(a)~Maximal values of the susceptibilities $\chi_4(t)$ and $\chi_{4\rm s}(t)$, plotted 
against the times $t_{\rm max}$ at which
these maximal values occur.  In contrast to Fig.~\ref{fig:chin-all}, these susceptibilities increase with $c$ (at fixed $T$).
(b) Comparison of $\Delta g_4(r,t)$ at the times that maximise $\chi_4(t)$, for $T=0.5$ and $c=0.00,0.04$ [these are the 
points with the largest and smallest values of $t_{\rm max}$ for $T=0.5$ in (a)].
The fits are of the form $\Delta g_4(r,t)=A\ee^{-r/\xi}$ with equal correlation lengths $\xi$ but different prefactors, 
illustrating that the variations in $\chi_4(t_\mathrm{max})$ with $c$ shown in panel (a) do not come from
an increasing length scale.  
}
\label{fig:chi4-all}
\end{figure}

In conclusion, we have presented measurements of four-point correlation functions in systems with
randomly pinned particles.  By carefully considering the normalisation and analysis of four-point functions,
we can extract dynamic length scales directly from real-space correlation functions.
At the moderate supercooling and pinning probabilities used, we do not find any evidence for an increasing
dynamic length scale, despite the significant increases in relaxation time.  If this behaviour persists
for large pinning and lower temperatures, it would be consistent with an inhomogeneous response of
random pinning, where different domains in the system stop relaxing at different values of $\pf$, and
there is a crossover from mobile to inactive states. However, 
we were not able (in these large systems) to approach very closely to any random pinning glass transition,
so we cannot rule out a crossover to the mean-field predictions of RFOT at larger pinning fractions and lower
temperatures.  This question remains one for further studies of both thermodynamic overlaps (as in~\cite{berthier-pin2013}) and
dynamical correlations, as in this work.

\begin{acknowledgments}
We thank Ludovic Berthier,  Giulio Biroli, Walter Kob, and Daniele Coslovich
 for helpful discussions related to random pinning, and the EPSRC
for funding through grant EP/I003797/1.
\end{acknowledgments}

\begin{appendix}

\begin{figure}
\includegraphics[width=6.5cm]{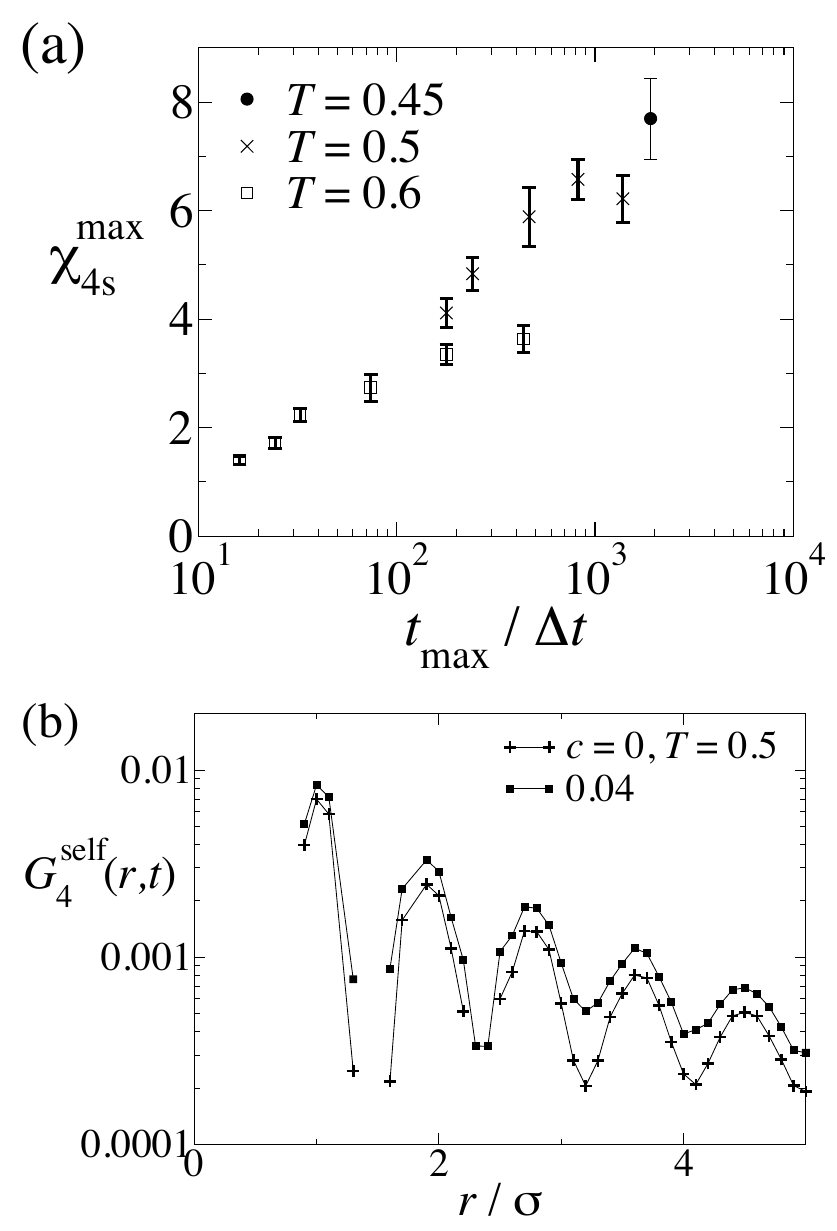}
\caption{
(a)~Maximal values of the self-part of the four-point susceptibility,
$\chi_{\rm 4s}(t)$.  (b)~Comparison of $G_4^\mathrm{self}(r,t)$, the spherical
average of $(\langle a^{\rm s}_i a^{\rm s}_j\rangle - \langle a^{\rm s}_i\rangle^2)/\langle n_i\rangle$, 
at the times that maximise $\chi_{\rm 4s}(t)$, with $T=0.5$ and $c=0.00,0.04$  [these are the 
points with the largest and smallest values of $t_{\rm max}$ for $T=0.5$ in (a)].  The susceptibility $\chi_{\rm 4s}$
is obtained by an integral of $G_4^\mathrm{self}(r,t)$: as in Fig.~\ref{fig:chi4-all}, the increase in susceptibility with $c$ comes from
an increasing prefactor (correlation amplitude) but not an increasing length scale.}
\label{fig:chis-all}
\end{figure}

\section{Variants of $\chi_4$}

As discussed in the main text, the main quantity of interest for dynamical heterogeneity
in these systems is the four-point correlation length $\xi_4$.
We have emphasised that we use a normalised four-point susceptibility $\chin(t)$, 
because of its transparent connection to $\xi_4$, via $g_4(r,t)$.  Here,
we present some extra numerical results using alternative definitions (or normalisations) of $\chi_4(t)$.

In Fig.~\ref{fig:chi4-all}(a), we show how the peak value of the conventional four-point susceptibility $\chi_4(t)$ depends on
$T$ and $c$, using a similar representation to Fig.~\ref{fig:chin-all}.  Unlike the normalised susceptibility
$\chin(t)$, we find that $\chi_4(t)$ does increase as pinning is introduced, at constant temperature.  However,
this increase is weaker than the dependence of $\chi_4(t)$ on temperature $T$.  Further, as discussed in the main
text, analysis of the four-point correlations in real space shows that the increase in $\chi_4(t)$ with pinning comes
primarily from an increase in the amplitude of four-point correlations, and not from an increasing length scale. 
This effect is illustrated in Fig.~\ref{fig:chi4-all}(b).

We also define a `self' variant of the four-point susceptibility.  That is, let $a^\mathrm{s}_i(t,t')$ be the number of particles
in cell $i$ at time $t$ that are also in the same cell at time $t'$.  Hence $Q_{\mathrm{s}}(t,t') = \sum_i a^\mathrm{s}_i(t,t')$
is the self-overlap and $\chi_{4\mathrm{s}}(t,t') = \frac{1}{N_\mathrm{A}} \langle \delta Q_\mathrm{s}(t,t')^2 \rangle$ the associated
susceptibility.  We show results in Fig.~\ref{fig:chis-all} for the peak value of $\chi_{4\mathrm{s}}(t,t')$ (we set $t'=0$ and suppress this argument, 
as above).  In the absence of pinning, we find
$\chi_{4\mathrm{s}}(t) \approx \chi_4(t)$ but in the presence of pinning then $\chi_{4\mathrm{s}}(t)$ is larger than the `collective'
$\chi_4(t)$.  From a physical point of view, $\chi_{4\mathrm{s}}(t)$ receives contributions
from motion where particles exchange their positions but preserve the overall profile; collective observables are blind to
such motion since total cell occupancies are unchanged.  While this effect seems interesting, analysis of the real-space correlations
indicates that the length scale associated with the self-correlations is very similar to that of the collective correlations shown in the main
text.  Thus, the difference in the correlation functions comes primarily through the amplitude of correlations [Fig.~\ref{fig:chis-all}(b)].  
For these `self'-correlations, the difficulty of finding
an appropriate normalisation for the amplitude of the $a^\mathrm{s}$-correlations
means that we are not able to ascribe any physical meaning to the decoupling of self and collective susceptibilities.  

Finally, we note that while our measurements used an ensemble where the number of pinned particles was free to fluctuate,
other simulation work has used a fixed number of pinned particles~\cite{berthier-pin2012,berthier-pin2013}.  
We note that while length scales
should be independent of the ensemble used, susceptibilities do differ~\cite{berthier-science05,berthier-chi4-07}: one has
\begin{equation}
\chi_4(t) = \chi_{4,\mathrm{fixed}}(t) + \left[\frac{\partial}{\partial c} \langle Q(t)\rangle\right]^2 \cdot \langle \delta c^2 \rangle \cdot \frac{1}{N_\mathrm{A}}
\end{equation}
where $\chi_{4,\mathrm{fixed}}(t)$ is the susceptibility evaluated with a fixed number of pinned particles,
and $\langle (\delta c)^2 \rangle=c(1-c)/N$ is the fluctuation in the concentration of pinned particles.  We note that
the second term can be significant in these systems, which again emphasises the care that must be taken
when interpreting measurements of four-point susceptibilities in these systems.

\end{appendix}

\end{document}